\documentclass[doublecol,figures]{epl2}

\usepackage{amssymb}
\usepackage{amsmath}
\usepackage{bm}

\newcommand{\bea}{\begin{eqnarray}}
\newcommand{\eea}{\end{eqnarray}}
\newcommand{\be}{\begin{equation}}
\newcommand{\ee}{\end{equation}}

\newcommand{\rt}[1]{{}}

\title{Phase structure and phase transitions
of the \mth{SU(2)\times O(N)} symmetric scalar field 
theory}
\shorttitle{Phase structure of the \mth{SU(2)\times O(N)} symmetric model} 

\author{
A. Patk{\'o}s \inst{1,2}\thanks{E-mail: \email{patkos@ludens.elte.hu}} 
\and 
Zs. Sz{\'e}p  \inst{3} \thanks{E-mail: \email{szepzs@achilles.elte.hu}}
}
\shortauthor{A. Patk{\'o}s and Zs. Sz{\'e}p}

\institute{
\inst{1}Department of Atomic Physics, E{\"o}tv{\"o}s University,
H-1117 Budapest, Hungary\\
\inst{2}Research Group for Statistical and Biological Physics of the
Hungarian Academy of Sciences, H-1117 Budapest\\
\inst{3}Research Institute for Solid State Physics and Optics of the
  Hungarian Academy of Sciences, H-1525 Budapest
}

\abstract{
Radiatively induced $SU(2)$ symmetry breaking is shown to be a genuine
feature of $SU(2)\times O(N)$ globally symmetric renormalisable field
theories in the large $N$ limit, describing interaction of a complex
$SU(2)$ doublet, $O(N)$-singlet field with an $SU(2)$ singlet, $O(N)$
vector. Symmetry breaking solutions are found even when all fields
have positive renormalised squared mass.  The emerging novel mechanism
of symmetry breaking can reproduce with a choice of $N\sim 300$ the
standard range of the electroweak condensate and the Higgs mass
occurring in the extended Higgs dynamics of an $SU(2)$ symmetric
Gauge+Higgs model.
}


\begin{document}

\maketitle

\section{Introduction}

\allowdisplaybreaks{
The Landau-Weiss mean field description of spontaneous symmetry breaking is
based on assuming negative sign to the quadratic term of the expansion of
the classical potential around the symmetric extremal point (mostly the
origin) of the order parameter field. It has been demonstrated by Coleman
and Weinberg, that for a classically conformal ($m^2=0$) theory renormalised
radiative corrections might generate such wrong sign mass term in the
effective potential \cite{coleman73}.

The present paper extends the validity of the Coleman-Weinberg phenomenon to
a large class of renormalisable models of interacting scalar fields all
having right (positive) sign squared masses. We shall study a complex
$SU(2)$ doublet $\Phi=(\sigma+i\phi_1,\phi_2+i\phi_3)$  which develops a 
nonzero vacuum expectation value $v$ under the influence of the fluctuations
of a multicomponent $SU(2)$ singlet field. We assume an $O(N)$ symmetric
quartic self-interaction for this latter field $\psi_i$, $i=1\dots N$ and an
$SU(2)\times O(N)$ symmetric interaction between the two fields resulting in
the  Lagrangian:
\begin{eqnarray}
L[\psi_i,\Phi]=\frac{1}{2}\partial_\mu\Phi^\dagger
\partial^\mu\Phi+\frac{1}{2}(\partial_\mu\psi_i)^2-\frac{1}{2}m_2^2\psi_i^2
\nonumber\\
-\frac{1}{2}m_3^2\Phi^\dagger\Phi
-\frac{\lambda_1}{24N}(\psi_i^2)^2-
\frac{\lambda_2}{24N}(\Phi^\dagger\Phi)^2
-\frac{\lambda_3}{12N}\psi_i^2\Phi^\dagger\Phi.
\label{Lagrangian}
\end{eqnarray}

The leading order large $N$ renormalised solution of this model was
constructed by us in a previous publication \cite{patkos06}.  This
model is a generalisation of the model of \cite{chivukula91}. There,
however, $\sigma$ and $\phi_j$, $j=1\dots 3$ real fields were part of
the $O(N)$ multiplet.  The clear distinction between the symmetries of
the $\Phi$ and $\psi_i$ fields leads to very different results. The
consequences of introducing a large(!) number ($N$) of $SU(2)$ singlet
scalars were discussed recently in \cite{bahat-treidel06}. These
authors apparently did not address the issue of $N$-scaling of the
couplings unavoidable to keep the self-energies ${\cal O}(N^0)$.

The model (\ref{Lagrangian}) can be interpreted as an extension of the
Higgs sector of the Standard Model, since after the symmetry breaking
the SM Higgs sector appears to have the same form as the $SU(2)$ part
of this model after the shift $\sigma\rightarrow\sigma+\sqrt{N}v$ (the
real field $\sigma$ corresponds to the Higgs particle, the other three
components are the would-be-Goldstone modes becoming the longitudinal
gauge excitations). Gauge fields interact with the scalar fields via
the covariant derivative of $\Phi$, and contribute to the equations of
both the vacuum condensate and the propagators of the Higgs fields. It
can be shown that in the leading large $N$ order this contribution is
subdominant in any gauge and to this order the symmetry breaking
effect is induced solely by the gauge singlet $O(N)$ vector. For this
reason the novel symmetry breaking mechanism to be presented here
works also for the Higgs effect in extensions of the Standard Model.

In Higgs physics at present renewed attempts are made to answer the
recurring question \cite{chivukula91,bahat-treidel06} ``Can Nature hide the
Higgs particle?'' With the advent of LHC experiments increasing variety of
alternative Higgs scenarios are proposed and analysed. Most extensions of
the standard Higgs sector ($\Phi$) by further scalars ($\psi$) start from
supersymmetry.  Consequences of adding singlet scalars in various versions
to minimal supersymmetric extensions are summarised in \cite{barger06}. 
Other singlet extensions were simply guided by the ``principle of minimal
modification'' 
\cite{patt06,calmet06,bahat-treidel06,davoudiasl05,bij06,o'connell06}.

The possibility of electroweak symmetry breaking induced by the vacuum
expectation value in the hidden (phantom) sector was first envisaged in
Ref.\cite{patt06}. One might note, however, that the tree level mechanism
proposed by these authors still assumes a destabilising negative sign for
the strength of the biquadratic Higgs-phantom interaction.  Radiative
symmetry breaking was also considered for such case by \cite{espinosa07}.
The extra scalars appear as natural candidates also for the role of the
cosmological inflaton and/or dark matter
\cite{davoudiasl05,ignatiev00,shaposhnikov06}.

The aim of this paper is to give a complete description the phase
structure of the model (\ref{Lagrangian}) for $N\rightarrow\infty$,
without relying on any weak coupling argument. Features of gauged
models will be commented where it is appropriate. The investigation
will focus on the region where all (quadratic, biquadratic and
quartic) couplings are {\it positive}. It will be shown that in this
case no simultaneous breakdown of the $O(N)$ and of the $SU(2)$
symmetry (e.g. $v\neq 0, u\neq 0$,~ cf. eq.(\ref{gen-sb})) is
possible. The statement remains valid even at next-to-leading order
(NLO) in the large $N$ expansion.

With an extensive numerical study it will be shown that in a
considerable part of the coupling space one finds solutions compatible
with our actual physical picture on the electroweak symmetry breaking.
The consistency of the solutions requires the cut-off
effects inherent for a trivial theory to lie above the physical
spectra and the normalisation scale. These effects are signalled in the
present renormalised formulation by the location of the Landau ghost
pole in the $\sigma$ channel. For moderate values of the quartic
couplings it turns out that generically $M_{\sigma} < 2m_\psi$,
therefore no hidden decays would hide the Higgs signal. Increasing
these couplings makes accessible consistent models with
$\sigma\to\psi\psi$ decay. The finite temperature fluctuations
destroying the vacuum expectation value of $\sigma$ are dominated in
this model by the hidden $O(N)$ multiplet. The range of the values of
$T_c$ follows that of the Higgs mass and is not very sensitive to the
quartic couplings.

\section{Leading large $N$ analysis of the phase structure at $T=0$}

The phase structure is investigated by applying to the Lagrangian density the 
shifts
\begin{equation}
\psi_1\rightarrow \psi_1+\sqrt{N}u,\qquad \sigma\rightarrow \sigma+\sqrt{N}v.
\label{gen-sb}
\end{equation}
The shifted Lagrangian is of the following form:
\begin{eqnarray}
\nonumber
L_\mathrm{free}&=&\frac{1}{2}(\partial_\mu\sigma)^2
+\frac{1}{2}(\partial_\mu\phi_j)^2+
\frac{1}{2}(\partial_\mu\psi_i)^2 \\\nonumber
&&-\frac{1}{2}m_2^2(\psi_i^2+2\sqrt{N}u\psi_1+Nu^2)\\\nonumber
&&-\frac{1}{2}m_3^2(\sigma^2+2\sqrt{N}v\sigma+Nv^2+\phi_j^2),\\\nonumber
L_\mathrm{int}&=&-\frac{\lambda_1}{24N}(\psi_i^2+2\sqrt{N}u\psi_1+Nu^2)^2\\
\nonumber
&&-\frac{\lambda_2}{24N}(\sigma^2+2\sqrt{N}v\sigma+Nv^2+\phi_j^2)^2
\nonumber\\
\nonumber
&&-\frac{\lambda_3}{12N}(\psi_i^2+2\sqrt{N}u\psi_1+Nu^2)\times
\\&&(\sigma^2+2\sqrt{N}v
\sigma+Nv^2+\phi_j^2).
\end{eqnarray}
The  equations for the expectation values of the fields $\psi_1,
\sigma$ are reached by taking first the appropriate functional
derivatives of the classical action:
\begin{eqnarray}
\nonumber
\frac{\delta S}{\delta\sigma}&=&
-(\Box+m_3^2+\frac{\lambda_2}{2}v^2+\frac{\lambda_3}{6}u^2)\sigma\\
\nonumber
&&-\sqrt{N}v(m_3^2+\frac{\lambda_2}{6}v^2+\frac{\lambda_3}{6}u^2)\\\nonumber
&&-\frac{\lambda_2}{6N}(\sigma^3+\phi_j^2\sigma)-\frac{\lambda_3}{6N}
\psi_i^2\sigma-
\frac{\lambda_2}{2\sqrt{N}}v\sigma^2\\
\nonumber
&&-\frac{\lambda_3}{6\sqrt{N}}v\psi_i^2-\frac{\lambda_3}
{3\sqrt{N}}u\psi_1\sigma-\frac{\lambda_3}{3}uv\psi_1,\\
\nonumber
\frac{\delta S}{\delta\psi_1}&=&-(\Box+m_2^2+\frac{\lambda_1}{2}u^2+
\frac{\lambda_3}{6}v^2)\psi_1\\
\nonumber
&&-\sqrt{N}u(m_2^2+\frac{\lambda_1}{6}u^2+\frac{\lambda_3}{6}v^2)
-\frac{\lambda_3}{3}uv\sigma\\
\nonumber
&&
-\frac{\lambda_1}{6N}\psi_i^2\psi_1-\frac{\lambda_3}{6N}\psi_1(\sigma^2+
\phi_j^2)-\frac{\lambda_3}{3\sqrt{N}}v\psi_1\sigma\\
&&-\frac{\lambda_1}{6\sqrt{N}}u
(3\psi_1^2+\psi_\alpha^2)-\frac{\lambda_3}{6\sqrt{N}}u(\sigma^2+\phi_j^2)
,
\end{eqnarray}
with $\alpha=2,...,N$.  The corresponding quantum equations are
obtained by replacing a generic field $\varphi_A$ by $\varphi_A+
G_{\varphi_A,\varphi_B}({\delta}/{\delta\varphi_B})$ on the right hand
sides of the previous equations cf. Eq.~(3) of \cite{patkos06}.

The leading order equations for the order parameters arise by keeping in
$\delta\Gamma/\delta\sigma$ and $\delta\Gamma/\delta\psi_1$ only terms
proportional to $\sqrt{N}$ and setting the fluctuating fields to zero:
\begin{equation}
\frac{\delta\Gamma}{\delta\sigma}\Big|_{\psi_i,\sigma=0}=
-\sqrt{N}v\left(m_3^2+\frac{\lambda_2}{6}v^2+
\frac{\lambda_3}{6}u^2+\frac{\lambda_3}{6}T_\psi\right)=0,
\label{equations-of-state1}
\end{equation}
\begin{equation}
\frac{\delta\Gamma}{\delta\psi_1}\Big|_{\psi_i,\sigma=0}=-\sqrt{N}u
\left(m_2^2+\frac{\lambda_1}{6}u^2+\frac{\lambda_3}{6}v^2+
\frac{\lambda_1}{6}T_\psi \right)=0.
\label{equations-of-state2}
\end{equation}
Here, in the tadpole contribution $T_\psi$ the propagator of the
$\psi_\alpha$ modes is understood. It has a well-known singular
$(T_{\psi,div})$ and a regular $(T_{\psi,F}$) part. 
$T_{\psi,F}$ (and also the bubble integral $I_{\psi,F}(M_\sigma)$
below) contains contributions inversely proportional to some power of
the cut-off. This cannot be sent to infinity because of the triviality
of the model. They can be neglected only if $m_\psi^2/\Lambda^2$ or
$M_\sigma^2/\Lambda^2$ are small \cite{kuti93,heller93}.  In
(\ref{equations-of-state1}) and (\ref{equations-of-state2}) the
tadpole contributions from $\sigma$ and $\phi_j$ are neglected as
${\cal O}(1/N)$ effects. The same refers to the contribution of the
gauge field tadpoles which couple to $\sigma$ through the covariant
derivative in an $SU(2)$ Gauge+Higgs model.

The renormalisation of the equations of state
(\ref{equations-of-state1}) and (\ref{equations-of-state2}) can be done with
help of the formulae established in the previous paper
\cite{patkos06}. For instance let us divide
(\ref{equations-of-state2}) by $\lambda_1$:
\begin{equation}
\sqrt{N}u\left(\frac{m_2^2}{\lambda_1}+\frac{1}{6}T_{\psi,\mathrm{div}}
+\frac{1}{6}u^2
+\frac{\lambda_3}{6\lambda_1}v^2+\frac{1}{6}T_{\psi,F}\right)=0.
\end{equation}
It renormalises by \cite{patkos06} to
\begin{equation}
\sqrt{N}u\left(\frac{m^2_{2,R}}{\lambda_{1,R}}+\frac{1}{6}u^2+
\frac{\lambda_{3,R}}{6\lambda_{1,R}}v^2+
\frac{1}{6}T_{\psi,F}\right)
=0.
\label{Eq:EoS-u}
\end{equation}
Let us multiply 
(\ref{equations-of-state1}) by $u$ and 
(\ref{equations-of-state2}) by $v$, and subtract
$\lambda_3/\lambda_1$ times the second from the first:
\begin{equation}
\sqrt{N}uv\left[m_3^2-\frac{\lambda_3}{\lambda_1}m_2^2+
\frac{1}{6}\left(\lambda_2-\frac{\lambda_3^2}{\lambda_1}
\right)v^2\right]=0,
\label{t-indep}
\end{equation}
which coincides with its renormalised form by the results of
\cite{patkos06}. These two renormalised equations imply also the
validity of the renormalised equation
\begin{equation}
\sqrt{N}v\left(m_{3,R}^2+
\frac{\lambda_{2,R}}{6}v^2+\frac{\lambda_{3,R}}{6}u^2+\frac{\lambda_{3,R}}{6}
T_{\psi,F}\right)=0.
\label{Eq:EoS-v}
\end{equation}

In what follows we will need also the gap-equation for the mass of the
$\psi_\alpha$ modes, obtained from the leading order
Dyson-Schwinger equation for $iG^{-1}_{\psi_\alpha}(p)= \delta^2
\Gamma/\delta\psi_\alpha\delta\psi_\alpha$ using the parametrisation
$iG^{-1}_{\psi_\alpha}(p)=p^2-m_\psi^2$:
\begin{equation}
m_\psi^2=m_{2,R}^2+\frac{\lambda_{1,R}}{6}u^2+
\frac{\lambda_{3,R}}{6}v^2+\frac{\lambda_{1,R}}{6}T_{\psi,F}.
\label{gap-psi}
\end{equation}
This equation is gauge independent by the $SU(2)$ singlet nature of
the $\psi_i$ field. Therefore the equations (\ref{equations-of-state1}),
(\ref{equations-of-state2}) and (\ref{gap-psi}) which determine the
phase structure to leading order in $N$ are gauge independent! 

The analysis of the phase structure starts by examining first the
existence of phases with partially broken symmetries (Case {\bf A} and
Case {\bf B} below). In these cases Eq.~(\ref{t-indep}) is satisfied
automatically, since it contains the factor $uv$.

\noindent
{\bf Case A}: Higgs condensate only ($u=0$, $v\neq 0$)

The relevant equation cf. (\ref{Eq:EoS-v}) is
\begin{equation}
m_{3,R}^2+
\frac{\lambda_{2,R}}{6}v^2+\frac{\lambda_{3,R}}{6}T_{\psi,F}=0.
\label{case-A-eos}
\end{equation}
The tadpole integral is determined by the solution of (\ref{gap-psi}).
This phase was qualitatively investigated in \cite{patkos06}. 

\noindent
{\bf Case B}: $O(N)$ symmetry breaking ($u\neq 0$, $v=0$)

The relevant equation of state cf. (\ref{Eq:EoS-u})  is the following:
\begin{equation}
m^2_{2,R}+\frac{\lambda_{1,R}}{6}u^2+\frac{\lambda_{1,R}}{6}T_{\psi,F}
=0.
\label{Eq:case-B-eos}
\end{equation}
The tadpole integral contains now the massless Goldstone
propagators. The fact that here the $\psi_\alpha$ components are
massless can be be readily checked by comparing
(\ref{Eq:case-B-eos}) and (\ref{gap-psi}).  This is the text-book case
of the large $N$ symmetry breaking in the $O(N)$ model, which
requires $m_{2,R}^2<0$.

\noindent
{\bf Case C}: Two-condensate phase

The interesting question of the possible existence of a
phase with two condensates
($u\neq 0, v\neq 0$) starts by noting that Eq.~(\ref{t-indep})
in this case implies a temperature independent $v$ condensate if the
renormalised couplings are chosen to ensure the positivity of
\begin{equation}
v^2=-6\frac{m_{3,R}^2-\lambda_{3,R}m_{2,R}^2/\lambda_{1,R}}
{\lambda_{2,R}-\lambda_{3,R}^2/\lambda_{1,R}}>0.
\label{c-v-condensate}
\end{equation}
Substituting this into any of the two original renormalised
equations of state, it will modify the renormalised mass term in a
kind of effective theory of the $\psi$ field:
\begin{equation}
\lambda_{3,R}\left(\frac{\lambda_{2,R}m_{2,R}^2-\lambda_{3,R}m_{3,R}^2}
{\lambda_{1,R}\lambda_{2,R}-\lambda_{3,R}^2}+\frac{1}{6}u^2
+\frac{1}{6}T_{\psi,F}\right)=0.
\label{c-u-condensate}
\end{equation}
Since the $O(N)$ symmetry is
broken Goldstone's theorem ensures the masslessness of the
propagators to be used in $T_{\psi,F}$. Due to this, at $T=0$ the
renormalised tadpole $T_{\psi,F}$ vanishes and the effective mass in
(\ref{c-u-condensate}) is unavoidably negative for a
physical $u$ condensate. From this and (\ref{c-v-condensate}) one
concludes that for the existence of phase {\bf C} the following two
inequalities should be satisfied:
\begin{equation}
\frac{\lambda_{2,R}m_{2,R}^2-\lambda_{3,R}m_{3,R}^2}
{\lambda_{1,R}\lambda_{2,R}-\lambda_{3,R}^2}<0,\quad
\frac{\lambda_{3,R}m_{2,R}^2-\lambda_{1,R}m_{3,R}^2}
{\lambda_{1,R}\lambda_{2,R}-\lambda_{3,R}^2}>0.
\end{equation}  
However, these inequalities cannot be satisfied simultaneously with
potentials whose all couplings are positive: $m_{i,R}^2>0$ and
$\lambda_{i,R}>0$. For the case of positive denominator the two
inequalities are satisfied only if
\begin{equation}
\frac{\lambda_{2,R}}{\lambda_{3,R}}<\frac{m_{3,R}^2}
{m_{2,R}^2}<\frac{\lambda_{3,R}}{\lambda_{1,R}}
\end{equation}
is valid for $\lambda_{1,R}\lambda_{2,R}>\lambda_{3,R}^2$. 
This requirement, however, contradicts the assumption on the
denominator. The same type of contradiction is arrived at when
$\lambda_{1,R}\lambda_{2,R}< \lambda_{3,R}^2$. The only way to
reconcile the two conditions is to choose both renormalised mass
squares to be negative: $m_{i,R}^2<0$.

It is tempting to guess next that the $v\neq 0, u\neq 0$ phase might
be of the form
\begin{equation}
v=v_0+\frac{1}{N}v_1,\qquad u=\frac{1}{N}u_1,
\label{NLO-ansatz}
\end{equation}
and Case {\bf C} might be realised as next-to-leading order
perturbation of Case {\bf A}. This can be checked by substituting this
Ansatz into the next-to-leading order (NLO) equations of state. The
equation for $u_1$ arises from $\delta\Gamma/\delta\psi_1$ which reads
with NLO accuracy as follows:
\begin{eqnarray}
\nonumber
\frac{\delta\Gamma}{\delta\psi_1}=-\sqrt{N}u\left(m_2^2+\frac{\lambda_1}{6}
u^2+\frac{\lambda_3}{6}v^2\right)-\\
\frac{\lambda_1}{6\sqrt{N}}u
(G_{\alpha\alpha}+3G_{11})-\frac{\lambda_3}{6\sqrt{N}}uG_{\sigma\sigma}
-\frac{\lambda_3}{3\sqrt{N}}vG_{1\sigma}-
\nonumber\\
\frac{i}{6N}\int_{A,B,C}
\left(\lambda_1G_{iA}G_{iB}G_{1C}+\lambda_{3}G_{\sigma A}
G_{\sigma B}G_{1C}\right)\Gamma^{(3)}_{ABC}.
\end{eqnarray}
The $\int_A$ symbols appearing above denote summation-integration over
discrete and continuous coordinates of the intermediary fields.  In
this equation only terms of ${\cal O}(1/\sqrt{N})$ are to be
kept. Therefore all propagators $G_{AB}$ are evaluated with leading
order accuracy only.  In view of the Ansatz (\ref{NLO-ansatz}) and the
fact that $G_{1\sigma}$ to leading order vanishes, one can neglect the
second, third and fourth terms on the right hand side of the
equation. In the last term the relevant 3-point functions are all
suppressed by at least a factor $1/\sqrt{N}$, therefore the conclusion
is that $u$ stays zero even at NLO.  This means that with all
couplings of the potential chosen positive in Case {\bf A} the
physically interesting 
Higgs (e.g. $SU(2)$) symmetry breaking phase
does not lead to mixing of the Higgs field and the hidden $O(N)$
multiplet.  

In the rest of the paper we concentrate on the analysis of 
Case {\bf A}.  We shall demonstrate at leading order in $N$ that it is
compatible in a large range of the parameters with our present
knowledge about the Higgs sector of the electroweak theory. 

\section{Leading order finite temperature behaviour of case A}

The analysis is based on combining Eq.~(\ref{case-A-eos}) with the gap
equation (\ref{gap-psi}) in which one sets $u=0$, and which selfconsistently
determines the mass of the $\psi$-field. When one eliminates
the renormalised tadpole integral $T_{\psi,F}$ from these two
equations one finds
\begin{equation}
\frac{1}{6}v^2=\frac{\lambda_{3,R}(m_{2,R}^2-m_\psi^2)-\lambda_{1,R}m_{3,R}^2}
{\lambda_{1,R}\lambda_{2,R}-\lambda_{3,R}^2}.
\label{order-param-psi}
\end{equation}
This equation can be reexpressed in terms of effective quantities which
are analogous to the expression of the vacuum expectation value in the
Standard Model:
\begin{eqnarray}
\nonumber
\frac{1}{6}v^2&=&-\frac{m_\mathrm{Higgs}^2}{\lambda_\mathrm{Higgs}N},\quad
\lambda_\mathrm{Higgs}N\equiv\lambda_{2,R}-
\frac{\lambda_{3,R}^2}{\lambda_{1,R}}\\
m_\mathrm{Higgs}^2&\equiv& m_{3,R}^2-\frac{\lambda_{3,R}}{\lambda_{1,R}}
(m_{2,R}^2-m_\psi^2),
\label{higgs-sm}
\end{eqnarray}
(recall that $\sqrt{N}v$ gives the physical strength of the
condensate). Stable symmetry breaking occurs when $m_\mathrm{Higgs}^2<0,$
$\lambda_\mathrm{Higgs}>0.$ It is clear that the temperature dependence of the
order parameter comes in only through $m_\psi^2$. Its equation arises
after eliminating $v^2$ from Eqs.(\ref{gap-psi}) and
(\ref{case-A-eos}):
\begin{equation}
\mu_\psi^2(1-C\ln(e\mu_\psi^2))=\mu_2^2-
\frac{\lambda_{3,R}}{\lambda_{2,R}}\mu_3^2+
\frac{\lambda_{1,R}\lambda_{2,R}-\lambda_{3,R}^2}{\lambda_{2,R}M_0^2}
T_\psi^{(T)},
\label{T-dep-gap}
\end{equation}
where the notations
\begin{equation}
\mu_i^2=\frac{m_{i,R}^2}{M_0^2},\quad
\mu_\psi^2=\frac{m_\psi^2}{M_0^2},
\quad
C=\frac{\lambda_{1,R}\lambda_{2,R}-\lambda_{3,R}^2}{96\pi^2\lambda_{2,R}}
\end{equation}
are introduced and the $T=0$ part of the tadpole integral is separated
from its temperature dependent part denoted by $T_\psi^{(T)}$.  The
explicit expression of the $T=0$ part is used on the left hand side of
(\ref{T-dep-gap}). Below we assume (cf. Eq.~(\ref{higgs-sm}))
 $\lambda_{1,R}\lambda_{2,R}-\lambda_{3,R}^2>0$,
that is $C>0$.

The left hand side of (\ref{T-dep-gap}) has a maximum for $C>0$ as a
function of $\mu_\psi^2$ at $\mu_{\psi,\mathrm{max}}^2=\exp(1/C)/e^2$.
Therefore, the condition for the existence of the solution of this
equation (at $T\neq0$) reads as
\begin{equation}
\frac{C}{e^2}e^{1/C}\geq
\mu_2^2-\frac{\lambda_{3,R}}{\lambda_{2,R}}\mu_3^2+\frac{\lambda_{1,R}
\lambda_{2,R}-\lambda_{3,R}^2}{\lambda_{2,R}M_0^2}T_{\psi}^{(T)}.
\end{equation}
Under this condition one finds two solutions for $\mu_\psi^2$, one is
smaller than $\mu_{\psi,\mathrm{max}}^2$ the other is larger. The smaller
starts to increase with $T$, which leads to decreasing value of
$v^2$. The phase transition is accessible to our treatment if the
above inequality is maintained until $v$ vanishes. The procedure which
determines $T_c$ starts by putting zero on the left hand side of
(\ref{order-param-psi}):
\begin{equation}
m_{\psi,c}^2=m_{2,R}^2-\frac{\lambda_{1,R}}{\lambda_{3,R}}m_{3,R}^2,
\end{equation}
where $m_{\psi,c}=m_\psi(T_c)$. This can be substituted into
Eq.~(\ref{case-A-eos}) in which we set $v=0$ and solve \rt{the gap
equation of $m_\psi$ which is solved} for \rt{finding} $T_c$:
\begin{eqnarray}
\nonumber
\frac{6m_{3,R}^2}{\lambda_{3,R}}
+T_{\psi,F}(m_{\psi,c}^2,T_c)
=\frac{6m_{3,R}^2}{\lambda_{3,R}}
+\frac{m_{\psi,c}^2}{16\pi^2}\ln\frac{em_{\psi,c}^2}{M_0^2}\\
+\frac{T_c^2}{2\pi^2}\int_{m_{\psi,c}/T_c}^\infty dx
\frac{\sqrt{x^2-m_{\psi,c}^2/T_c^2}}{e^x-1}=0.
\label{Eq:Tc}
\end{eqnarray}
 
When one starts to increase the temperature the right hand side of
Eq.~(\ref{T-dep-gap}) first increases at all values of $m_\psi$. As a
consequence it cuts the temperature independent left hand side in such
a way that both roots move towards $m_{\psi,\mathrm{max}}$. The
temperature value when the roots become degenerate is the maximal one
for which the solution of our model has sensible physics. If this
happens before $v$ reaches zero, one can not access the phase
transition. Depending on the actual couplings it might happen that the
two poles become degenerate at a mass value larger than
$m_{\psi,\mathrm{max}}$.  In this case Eq.~(\ref{Eq:Tc}) will have a
solution even when $m_{\psi,c}>m_{\psi,\mathrm{max}}$.  A particular
set of renormalised parameters is considered by us acceptable if the
Higgs transition falls into the temperature range of validity of the
model.

\section{Numerical study of case A}

We restrict our further investigation to the special case:
$\lambda_{2,R}=\lambda_{3,R}= \lambda,$
$\lambda_{1,R}=\lambda+\lambda'$, in order to diminish the number of
the tunable parameters. The results presented below should be
representative for the more general parameter choice.

The renormalised parameters of the theory are fixed using the $T=0$
equation of state for $v$ and the gap equation of $m_\psi$. In
addition one can make use of the gap equation of $\sigma$, obtained to
leading order in $N$ by summing up the insertion of bubbles $I_\psi$
in the $\sigma$ self-energy \cite{patkos06}.  In the gauged theory one
should add to the Higgs self-energy the contribution of the one loop
gauge diagrams. Here we omit this contribution. This way the following
set of equations is to be solved:
\begin{eqnarray}
&& v^2+\frac{6m_{3,R}^2}{\lambda}=-\frac{m_\psi^2}{16\pi^2}\ln\frac{em_\psi^2}
{M_0^2}\leq\frac{M_0^2}{16\pi^2e^2} ,
\label{reduced-eqs1}
\\
&&M_\sigma^2=\frac{v^2}{3}\frac{\lambda}{\lambda+\lambda'}\left[\lambda'+\frac
{\lambda}{1-\frac{\lambda+\lambda'}{6}I_{\psi,F}(M_\sigma^2)}\right],
\label{reduced-eqs2}
\\
&&m_{2,R}^2=m_\psi^2+m_{3,R}^2-\frac{\lambda'}{96\pi^2}m_\psi^2\ln
\frac{e m_\psi^2}{M_0^2}.
\label{reduced-eqs3}
\end{eqnarray}
The upper bound on $v$ displayed in Eq.~(\ref{reduced-eqs1}) restricts
it to values which are at least $4\pi e$ smaller than the
renormalisation scale. In order to have the physical value for the
electroweak condensate $\sqrt{N}v\approx 250\un{GeV}$ with a
normalisation scale $M_0$ at least twice as large, one has to choose
$N>(2\pi e)^2\simeq 292$.  In this range our leading large $N$
solution should work quite well. The compatibility of this large
number of ``dark'' degrees of freedom with cosmological constraints
should be investigated.

Our strategy for investigating the solutions of these equations is to
fix a reasonable value for the Higgs mass and some acceptable value
for the renormalisation scale $M_0$. The latter is chosen below the
unitarity limit of validity of the scalar theory.  Since at present
direct search results combined with electroweak precision tests
indicate $114\un{GeV}\leq M_H\leq 200\un{GeV}$, below we choose
$M_\sigma=140\un{GeV}$, $M_0=500\un{GeV}$ (and also
$M_0=800\un{GeV}$). Then $\lambda$ is varied in the range $\lambda\in
(0,400)$. $v^2$ and $m_{3,R}^2$ are chosen to respect the inequality
appearing in (\ref{reduced-eqs1}).

Next, the first equation of the above set allows to find $m_\psi$.
This equation has two solutions like Eq.~(\ref{T-dep-gap}), one root
is below $M_0/e$ the other one is above it.  Using one of the roots in
the renormalised bubble integral $I_{\psi,F}(M_\sigma)$, one finds
$\lambda'$ from (\ref{reduced-eqs2}) which can be rewritten in a very
enlightening way:
\begin{equation}
 \lambda'=\frac{6}{I_{\psi,F}(M_\sigma)}-\left(\frac{1}{\lambda}-
\frac{v^2}{3M_\sigma^2}\right)^{-1}.
\label{lambda-prime}
\end{equation}
Finally $m_{2,R}^2$ is determined from the third equation of the above
set, Eq.~(\ref{reduced-eqs3}). These two couplings depend on which
$m_\psi$ root was chosen.


A very restrictive criterion used to select the allowed models
is represented by the choice of the Landau ghost's scale defined as
the absolute value of the imaginary pole solution of the Higgs
propagator. Its equation is the imaginary continuation of
(\ref{reduced-eqs2}) with the following analytic form of the bubble
integral for $M_\sigma=im_L$:
\begin{equation}
I_{\psi,F}(m_L)=\frac{1}{16\pi^2}\ln\left[\frac{m_\psi^2}{M_0^2}
\left(\frac{Q-1}{Q+1}\right)^{-Q}\right],
\end{equation}
where $Q=\sqrt{1+\frac{4m_\psi^2}{m_L^2}}$. Only those parameter sets
are accepted where the Landau-ghost scale is above $M_0$. 
This criterion is dictated by the expectation that a
solution which fulfills the relations $M_\sigma,m_\psi \ll M_0<\Lambda
<m_L$ coincides with that of a fixed cut-off analysis 
carried out in the spirit of Refs.~\cite{kuti93,heller93}.

Using the above criterion practically all points are excluded in the
$\lambda$ region $(0,200)$ while for $\lambda\in(200,400)$ one finds
acceptable sets which allow us to densely populate the two branches of
$m_\psi$ roots. It turns out that the $m_\psi$ values obtained from
(\ref{reduced-eqs1}) fall on the physical branch of (\ref{T-dep-gap}),
that is for which $v$ decreases with increasing $T$. Eventually the
allowed parameters lead to $\lambda_\mathrm{Higgs}$ of
Eq.~(\ref{higgs-sm}) in the range $(0,0.9)$.

It is worth mentioning that the 3-point function
$\Gamma_{\psi_\alpha\psi_\alpha\sigma}$ and 4-point function
$\Gamma_{\psi_\alpha\psi_\alpha\psi_\alpha\psi_\alpha}$ taken at
vanishing momentum, which are the coefficients up to a negative sign
of the cubic and quartic values in the leading large $N$ expression of
the effective potential $V_\mathrm{eff}(\psi,\sigma)$, are actually
negative only in the range where the ghost scale is above
$M_0$. This ensures the stability of the effective potential.

\begin{figure}
\onefigure[width=0.49\textwidth]{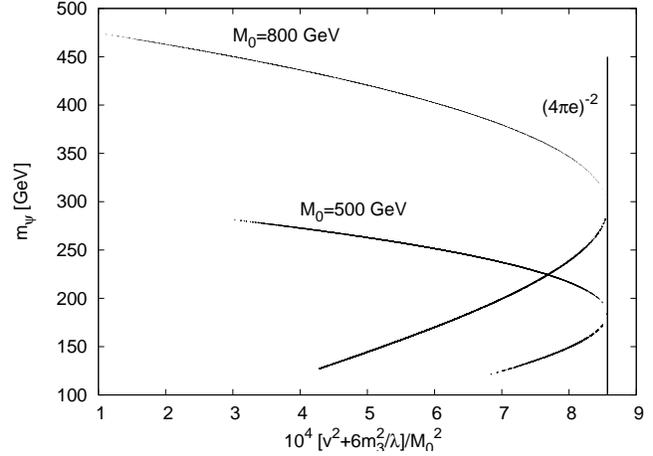}
\caption{The solution of Eq.~(\ref{reduced-eqs1}) for two different
  values of the normalisation scale $M_0$.}
\label{Fig1}
\end{figure}

\begin{figure*}[t]
\begin{center}
\hspace*{-0.5cm}
\includegraphics[keepaspectratio, width=0.55\textwidth,angle=0]
{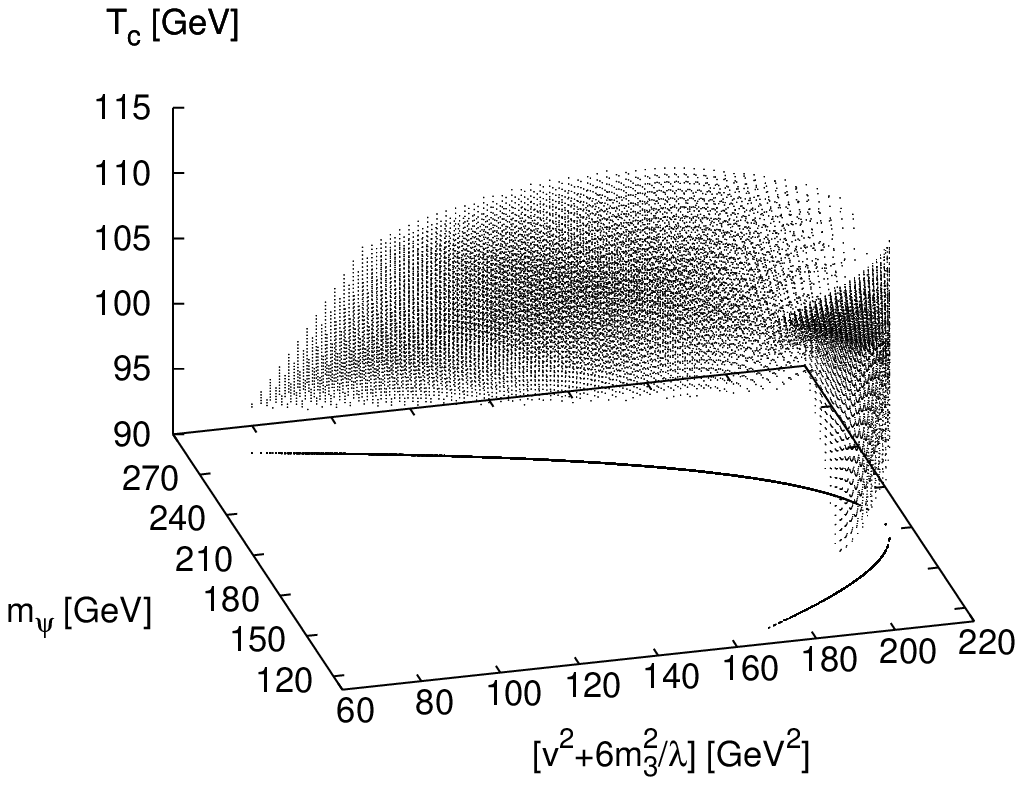}
\hspace*{-1.75cm}
\includegraphics[keepaspectratio, width=0.55\textwidth,angle=0]
{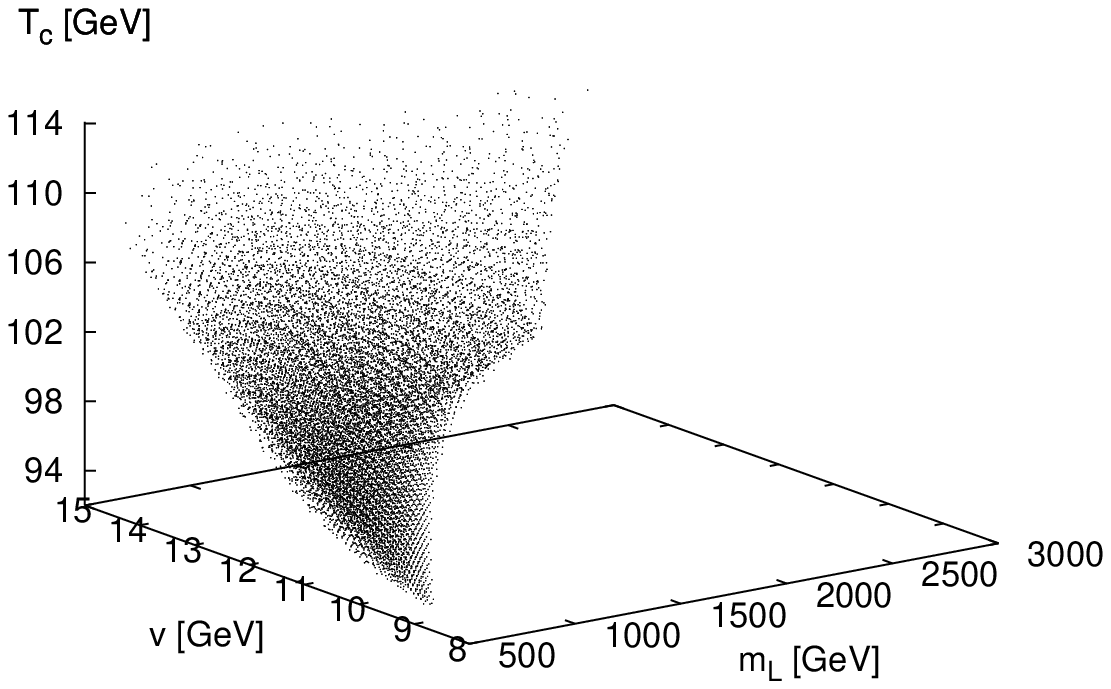}
\end{center}
\caption{The dependence of $T_c$ on an appropriate combination of
coupling parameters and/or physical data of the model. The lines in
the horizontal plane of the figure on the left hand side display the
two branches of the $m_\psi$ curve of Fig.~\ref{Fig1}. 
The figure on the right hand side presents $T_c$ as function
of $v$ and $m_L$ for the upper branch of $m_\psi$.}
\label{Fig2}
\end{figure*}
In Fig.~\ref{Fig1} we show the two branches of $m_\psi$ for $M_0=500,
800\un{GeV}$ as found by solving (\ref{reduced-eqs1}).  The shorter lower
arm is the consequence of the cut applied through the implementation
of the ghost criterion. The range of the $m_\psi$ values is
mostly above the Higgs mass (e.g. $140\un{GeV}$ in the present example).

The temperature of the phase transition falls 
slightly below this range. The spreading of $T_c$ over the two
branches is displayed in the left hand side of Fig.~\ref{Fig2} for
$M_0=500\un{GeV}$. A well-defined region, sharply limited both from above
and from below is filled by the values of $T_c$ found for the
different acceptable models. It is interesting to see that the region
filled by the $T_c$ values shows a certain characteristic shape when
displayed as a function of $v$ and $m_L$, the Landau-ghost scale.
This can be seen for the upper branch $m_\psi$ in the right hand side
of Fig.~\ref{Fig2}. The lower edge of the $T_c$ surface is due to the
requirement of not allowing models with much too low Landau-ghost
scale (low cut-off value).

\section{Conclusions}

In this paper we have pointed out the possibility of having
Coleman-Weinberg type symmetry breaking for an order parameter field
with positive renormalised mass parameter, induced by quantum
fluctuations of a multicomponent hidden ``phantom'' field.  Numerical
investigation showed that one finds large sets of positive
renormalised couplings of the extended model which lead to
$m_\mathrm{Higgs}^2<0, \lambda_\mathrm{Higgs}>0$ (e.g.
(\ref{higgs-sm})). The wrong-sign mass terms receives in
this way a natural origin.

For $\lambda\in(0,400)$ and $M_0=500\un{GeV}$ the solution of the gap
equation of the phantom field always leads to heavy quanta:
$2m_\psi>M_{\sigma}$, and also the Higgs-phantom mixing was shown to
be absent even at NLO level. Therefore in this toy model (where the
gauge field contribution to the Higgs mass is omitted) we conclude
that in a large part of the parameter space one would discover the
Higgs particle with the characteristics predicted by the Standard
Model. Scanning through the solutions obtained in a wider range of the
$\lambda$ we find that also lower $m_\psi$ values become accessible as
$\lambda$ grows. Above $\lambda\approx 750$ solutions appear for which
$m_\psi<M_\sigma/2$. The corresponding range of
$\lambda_\mathrm{Higgs}$ is $(0.8,0.94)$. Also, the critical
temperatures overlap with the values found for models with no hidden
Higgs decay.

More realistic investigations ought to include also gauge and fermion
contributions to the Higgs self-energy. The order of the finite temperature
phase transition can be examined conveniently by combining large $N$
techniques with finite temperature dimensional reduction.

\acknowledgements

This research was supported by the Hungarian Scientific Research Fund (OTKA)
under the grants No. T-046129 and PD-050015. Fruitful discussions
with J. Kuti and A. Jakov\'ac are acknowledged.

}

\end{document}